\documentclass{article}
\usepackage{amssymb,amsmath,cite,bm,graphicx}
\def\rd{\mathrm{d}}
\def\ri{\mathrm{i}}
\def\Dm{\mathrm{D}}
\def\dep{\delta_\mathrm{p}}
\begin{document}
\begin{titlepage}
\begin{center}
{\large \textbf{
The similarity of attractive and repulsive forces on a lattice}}

\vspace*{2\baselineskip}

Marjan-S. Mirahmadi, 
Amir~H.~Fatollahi\footnote{fath@alzahra.ac.ir}, \&
Mohammad Khorrami\footnote{mamwad@mailaps.org}
\\
\vspace{\baselineskip}
\textit{ Department of Physics, Alzahra University, Tehran
1993891167, Iran }
\end{center}
\vspace{\baselineskip}
\begin{abstract}
\noindent On a lattice, as the momentum space is compact,
the kinetic energy is bounded not only from below but also from above.
It is shown that this, somehow removes the distinction between
repulsive and attractive forces. In particular, it is seen that
a region with attractive force would appear forbidden for states
with energies higher than a certain value,
while repulsive forces could develop bound-states. An explicit transformation
is introduced which transforms the spectrum of a system corresponding
to a repulsive force, to that of a similar system corresponding to
an attractive force. Explicit numerical examples are presented for discrete energies
of bound-states of a particle experiencing repulsive force by a piecewise constant
potential. Finally, the parameters of a specific one dimensional translationally invariant system
on continuum are tuned so that the energy of the system resembles
the kinetic energy of a system on a one dimensional lattice. In particular,
it is shown that the parameters could be tuned so that while the width of
the first energy band and its position are kept finite, the energy gap
between the first energy band and the next energy band go to infinity,
so that effectively only the first energy band is relevant.
\end{abstract}

\vspace{2\baselineskip}


\end{titlepage}

\section{Introduction}
Formulation of physics on lattices has a distinguished place
in many areas of physics, from condensed matter physics to high energy physics.
In condensed matter physics, lattice formulations are mainly used
as an approximation to the real model, which is on the continuum.
For example as a crystal is symmetric under discrete translations,
to study the motion of a particle one could use a simple lattice
model (rather than the continuum) which has the same symmetries.
Such a simplified model does lead to a band structure for the energy spectrum,
although it results in only a single band, \cite{Saku}.

In high energy physics, there are cases where lattices
are used as approximations of continuum, examples of which are
lattice field theories and specially lattice gauge theories as
numerical approximations for the original theories on continuum
\cite{wilson, smith}. There are, however, models which regard the space
to be fundamentally discrete. Examples are theories formulated on
noncommutative spaces of Lie type noncommutativity, where the underlying
Lie group is compact \cite{jablam, su2nc}.

A remarkable feature of mechanics on a lattice is that the corresponding
momentum space is compact. So for a particle on a lattice, any continuous
function of the momentum, including kinetic energy, would be bounded
both from below and above.

The purpose of the present work is to explore the consequences of
the above-mentioned feature. In particular, it is shown that as a result of
the kinetic energy being bound from both below and above, the distinction
between repulsive and attractive forces is somehow blurred. As a result,
a region with positive kinetic energy, which is supposed to be allowed
would be forbidden if the energy is so high that the kinetic energy exceeds its
upper bound. Similarly, systems with repulsive forces could develop bound-states.
As will be seen, there is in fact a one-to-one correspondence
between the spectrum (bounded or unbounded) of a system with a certain
potential energy, and that of another system with a potential energy minus
that of the former.

To have a taste of the exact statement, let us consider the cases
when potential at infinity, $U(\infty)$, is larger (smaller) than
the potential at finite distances, corresponding to attractive
(repulsive) forces. For a particle on a continuum, the total energy
cannot be everywhere smaller than the potential energy. Hence for repulsive
forces the energy is larger than $U(\infty)$, meaning that far distances are
in classically allowed regions, and the system is unbounded.
Furthermore, there is no constraint on the energy levels, except that they should
be larger than $U(\infty)$. Hence the energy levels are unbounded from above.
For attractive forces, there could be energy levels which are smaller than
$U(\infty)$. These correspond to bound-states, where far distances are in
the classically forbidden region.

On a lattice, however, the kinetic energy is bounded from both below and above.
That results in the classically forbidden region to consist of points
where the kinetic energy is either less than its lower bound ($K_\mathrm{min}$), or
larger than its upper bound ($K_\mathrm{max}$). Hence, the energy cannot be
everywhere smaller than the potential energy plus $K_\mathrm{min}$ or larger than
the potential energy plus $K_\mathrm{max}$. An energy level which is larger than
$U(\infty)+K_\mathrm{max}$, corresponds to a bound-state. So repulsive forces
could result in bound states as well. Also, for attractive forces, there is
no energy level larger than $U(\infty)+K_\mathrm{max}$.

The scheme of the paper is the following. In section 2, the relation of
the position space being discrete (lattice) with the momentum space
being compact is investigated, with particular attention on the kinetic
energy as a continuous function of the momentum space. In section 3,
a system is studied which consists of a particle on a one dimensional
lattice, experiencing a position dependent potential energy.
The corresponding eigenvalue problem is discussed, including the relation
between the spectrums corresponding to attractive and repulsive forces,
and a particular example is studied in detail, to find the number and
values of the bound energy levels. In section 4, a toy model is
presented in which a particle moves on continuum under the influence of
a periodic potential energy. The model is a Kronig-Penney type \cite{Gasi},
consisting a particle moving in a periodic array of Dirac delta
potential energies. It is shown that the parameters of the model
could be tuned so that of the width and position the first energy band
is kept fixed, while the gap between the first band and the next band tend to
infinity. The first band could then be regarded as the values of
the kinetic energy for a particle moving on a lattice. So a system on
the continuum could be effectively changed to a system on a lattice.
\section{Lattice and the compact momentum space}
Consider a one dimensional infinite lattice. The Hilbert space for a particle
living on such a lattice is spanned by the set of orthonormal kets $|n\rangle$
(with $n$ being integer):
\begin{equation}\label{0.1}
\langle n|n'\rangle=\delta_{n\,n'}.
\end{equation}
So $|\psi\rangle$, the state vector of the particle can be represented by
the wave-function $\psi$ with
\begin{equation}\label{0.2}
\psi(n)=\langle n|\psi\rangle.
\end{equation}
Alternatively, one can use the Fourier-transformed basis, the set of
the orthonormal kets $|s\rangle$:
\begin{align}\label{0.3}
|s\rangle&=\frac{1}{\sqrt{2\,\pi}}\,\sum_{n=-\infty}^\infty \exp(-\ri\,n\,s)\,|n\rangle.\\ \label{0.4}
|n\rangle&=\int_{-\pi}^\pi\frac{\rd s}{\sqrt{2\,\pi}}\,\exp(\ri\,n\,s)\,|s\rangle.
\end{align}
It is seen that
\begin{equation}\label{0.5}
\langle s|s'\rangle=\dep(s-s'),
\end{equation}
where $\dep$ is the periodic Dirac delta:
\begin{equation}\label{0.6}
\dep(s)=\sum_{j=-\infty}^\infty\delta(s-2\,\pi\,j).
\end{equation}

The Kinetic energy $K$ would be a function of the lattice translation $T$ with
\begin{equation}\label{0.7}
T\,|n\rangle=|n+1\rangle.
\end{equation}
It is seen directly from (\ref{0.3}) that $|s\rangle$ is an eigenket of $T$:
\begin{equation}\label{0.8}
T\,|s\rangle=\exp(\ri\,s)\,|s\rangle.
\end{equation}
As the kinetic energy is a function of $T$, the ket $|s\rangle$ should be
an eigenket of $K$ as well:
\begin{equation}\label{0.9}
K\,|s\rangle=\tilde K(s)\,|s\rangle,
\end{equation}
where $\tilde K(s)$ is the corresponding eigenvalue. As
\begin{equation}\label{0.10}
|s+2\,\pi\rangle=|s\rangle,
\end{equation}
the set of the values of $s$ resulting in distinct kets $|s\rangle$
is compact (the momentum space is compact). Hence (assuming that the function
$\tilde K$ is continuous), the set of the eigenvalues of the kinetic energy
is also compact, namely it has a minimum $K_\mathrm{min}$ and a maximum
$K_\mathrm{max}$. The function $\tilde K$ should also be periodic
(with the period $2\,\pi$) in $s$. In the case of free particle, for which
the Hamiltonian $H$ has only the kinetic part, the energy is bounded from both
below (similar to the case of a free particle on continuum) and above.

A simple choice for the Kinetic energy is
\begin{equation}\label{0.11}
\tilde K(s)=K_0\,(1-\cos s).
\end{equation}
This is periodic, behaves like $s^2$ for small values of $s$, and
is increasing with $|s|$ as long as $s$ is in $[-\pi,\pi]$. One can express
the constant $K_0$ in terms of the dimensional parameters $m$ (the mass
of the particle) and $\ell$ (the lattice spacing):
\begin{equation}\label{0.12}
K_0=\frac{\hbar^2}{m\,\ell^2}.
\end{equation}
For the simple choice (\ref{0.11}), one has
\begin{align}\label{0.13}
K&=\frac{K_0}{2}\,(2-T-T^{-1}),\nonumber\\
&=-\frac{K_0}{2}\,\Dm^2,
\end{align}
where $\Dm^2$ is the Laplacian on the lattice:
\begin{equation}\label{0.14}
(\Dm^2\,\psi)(n)=\psi(n+1)+\psi(n-1)-2\,\psi(n).
\end{equation}
Obviously, for the kinetic energy of the form (\ref{0.11}),
\begin{align}\label{0.15}
K_\mathrm{min}&=0.\nonumber\\
K_\mathrm{max}&=2\,K_0.
\end{align}
\section{Position-dependent forces}
Similar to the case of a particle moving on a one dimensional (continuous line),
the Hamiltonian $H$ of a particle moving influenced by a position-dependent force
on a lattice, is the sum of the kinetic energy and the potential energy $U$,
\begin{equation}\label{0.16}
H=K+U,
\end{equation}
where $U$ is a function of the lattice site. That is, $|n\rangle$
is an eigenket of $U$ with the eigenvalue $U(n)$:
\begin{equation}\label{0.17}
U\,|n\rangle=U(n)\,|n\rangle.
\end{equation}
Corresponding to any eigenvalue $E$ of the Hamiltonian $H$,
the (classically) forbidden region is defined as the set of points $n$ for which
\begin{equation}\label{0.18}
[E-U(n)]\notin[K_\mathrm{min},K_\mathrm{max}].
\end{equation}
As the eigenvalues of the kinetic energy belong to the set $[K_\mathrm{min},K_\mathrm{max}]$,
the expectation value of $(K-K_\mathrm{min})\,(K_\mathrm{max}-K)$ is nonnegative.
That is, for any ket $|\psi\rangle$,
\begin{equation}\label{0.19}
\langle\psi|\,(K-K_\mathrm{min})\,(K_\mathrm{max}-K)\,|\psi\rangle\geq 0.
\end{equation}
So,
\begin{equation}\label{0.20}
\langle\psi|\,(H-U-K_\mathrm{min})\,(K_\mathrm{max}-H+U)\,|\psi\rangle\geq 0.
\end{equation}
This is true if $|\psi\rangle$ is an eigenket of $H$ with corresponding to the eigenvalue
$E$ as well. In that case, one arrives at
\begin{equation}\label{0.21}
\langle\psi|\,(E-U-K_\mathrm{min})\,(K_\mathrm{max}-E+U)\,|\psi\rangle\geq 0,
\end{equation}
or
\begin{equation}\label{0.22}
\sum_n[E-U(n)-K_\mathrm{min}]\,[K_\mathrm{max}-E+U(n)]\,|\psi(n)|^2\geq 0.
\end{equation}
The multiplier of $|\psi(n)|^2$ in the above is negative, if $n$ belongs to
the forbidden region. This shows that the Hamiltonian has no eigenvalue for
which everywhere is in the forbidden region. This is a generalization of
the statement for the continuum, that there Hamiltonian has no eigenvalue
which is less than the minimum of the potential energy (if such a minimum
exits).
\subsection{Relation between attractive and repulsive forces}
Consider a Hamiltonian of the form (\ref{0.16}). Let us restrict
to the case that the kinetic energy has the following property
\begin{equation}\label{0.23}
\tilde K(s+\pi)+\tilde K(s)=K_\mathrm{max}-K_\mathrm{min}.
\end{equation}
This is obviously satisfied by the simple choice (\ref{0.11}). In fact
it mean that $\tilde K$ is a constant plus an odd function of
$\exp(\ri\,s)$. Defining $H'$
similarly but with $(-U)$ instead of $U$, it is seen that
\begin{align}\label{0.24}
H'&=K-U,\nonumber\\
&=-[(K_\mathrm{max}-K_\mathrm{min}-K)+U]+(K_\mathrm{max}-K_\mathrm{min}).
\end{align}
Defining the unitary operator $S$ through
\begin{equation}\label{0.25}
S\,|n\rangle:=\exp(\ri\,\pi\,n)\,|n\rangle,
\end{equation}
it is seen that
\begin{equation}\label{0.26}
\{S,T\}=0,
\end{equation}
hence
\begin{equation}\label{0.27}
S\,K\,S^{-1}=K_\mathrm{max}-K_\mathrm{min}-K.
\end{equation}
It is also seen that
\begin{equation}\label{0.28}
[S,U]=0.
\end{equation}
So,
\begin{equation}\label{0.29}
H'=-S\,H\,S^{-1}+K_\mathrm{max}-K_\mathrm{min}.
\end{equation}
This shows that the eigenvalues and eigen-functions of $H'$
have a simple relation to those of $H$, namely if
\begin{equation}\label{0.30}
H\,|\psi\rangle=E\,|\psi\rangle,
\end{equation}
then
\begin{equation}\label{0.31}
H'\,|\psi'\rangle=E'\,|\psi'\rangle,
\end{equation}
(and, of course, vice versa) where
\begin{align}\label{0.32}
|\psi'\rangle&=S\,|\psi\rangle.\\ \label{0.33}
E'&=-E+K_\mathrm{max}-K_\mathrm{min}.
\end{align}
One has
\begin{equation}\label{0.34}
\psi'(n)=\exp(\ri\,\pi\,n)\,\psi(n).
\end{equation}
If $U$ corresponds to a so called attractive force, then
$(-U)$ corresponds to a repulsive force. But the spectrum of $H'$
is similar to that of $H$. Especially, if the eigenvector $|\psi\rangle$
of $H$ is a bound-state, that is the corresponding wave-function decays
exponentially when its argument tends to $\pm\infty$, then so is
the corresponding eigenvector $|\psi'\rangle$ of $H'$. Here, repulsive
forces do give rise to bound-states as well, in contrast to what happens in
continuum. But the bound-states corresponding to the repulsive forces occur
for high energies, rather than low energies, where the bound-states corresponding
to attractive forces occur.
\subsection{Piecewise constant potential energies}
When the potential energy is a piecewise constant function of
the (discrete) position, the eigenket of the Hamiltonian is found
similar to the case of the continuum, namely, one takes
the eigenket $\psi$ to be piecewise a linear combination of
plane waves. In each region (where the potential energy is constant)
the wave number is real if the region is not in the forbidden region.
Otherwise, the wave numbers are complex. To solve the eigenvalue problem,
one should write the continuity equation at boundaries.

For definiteness, for now on it is assumed that the kinetic energy is of
the form (\ref{0.11}). Then in each region, the eigenket $\psi$ is
a linear combination of two plane waves. At each boundary, one writes
a continuity condition and the eigenvalue equation corresponding to
that point. So if there are $q$ regions, one has $2\,(q-1)$
equations between the coefficients of the wave-function. In each of
the two regions which extend to $-\infty$ and $+\infty$, if the
wave number is not real the boundary condition that the wave-function does not
blow up at infinity makes the coefficient of one of the plane waves
vanish. So there are $2\,q$ coefficients of the plane waves, if
neither of the regions extending to infinity are forbidden,
$(2\,q-1)$ coefficients of the plane waves, if one and only one
of the regions extending to infinity are forbidden, and $(2\,q-2)$
coefficients of the plane waves, if both of the regions extending to infinity
are forbidden. Then, similar to the case of the continuum,
the spectrum of the Hamiltonian is continuous and the eigenspace is
two-dimensional if neither of the regions extending to infinity is
forbidden, the spectrum of the Hamiltonian is continuous and the eigenspace
is one-dimensional if one and only one of the regions extending to infinity
is forbidden, and the spectrum of the Hamiltonian is discrete if both of
the regions extending to infinity are forbidden. The difference with
the case of the continuum is that here the forbidden region is where
the energy is either smaller than the potential energy or larger than
the potential energy plus $2\,K_0$.

An example of a piecewise constant potential energy is
\begin{equation}\label{0.35}
U(n)=\begin{cases}+\infty,& n\leq0\\
U_\mathrm{I},& 0< n\leq n_0\\
U_\mathrm{II},&n>n_0
\end{cases}.
\end{equation}
The energy eigenket $\psi$ corresponding to such a potential
energy is found in a way similar to the case of continuous space, namely
\begin{equation}\label{0.36}
\psi(n)=\begin{cases}0,& n\leq0\\
\psi_\mathrm{I},& 0< n\leq n_0\\
\psi_\mathrm{II},&n>n_0
\end{cases}.
\end{equation}
Denoting the regions $0<n<n_0$ and $n>n_0$ with I and II, respectively,
it is seen that the eigenvalue equation in these regions would be
\begin{equation}\label{0.37}
(E-U_a)\,\psi_a=-\frac{K_0}{2}\,\Dm^2\psi_a,
\end{equation}
where $a$ is either I or II.
The solution to (\ref{0.37}) is
\begin{equation}\label{0.38}
\psi_a(n)=A_a\,\exp(\ri\,s_a\,n)+B_a\,\exp(-\ri\,s_a\,n ),
\end{equation}
with $A_a$ and $B_a$ being constants, and $s_a$ determined through
\begin{equation}\label{0.39}
E-U_a=K_0\,(1-\cos s_a).
\end{equation}
The boundary condition at $n=0$ reads
\begin{equation}\label{0.40}
A_\mathrm{I}+B_\mathrm{I}=0,
\end{equation}
so that
\begin{equation}\label{0.41}
\psi_\mathrm{I}(n)=A\,\sin(s_\mathrm{I}\,n).
\end{equation}
The boundary conditions at $n=n_0$ are
\begin{align}\label{0.42}
A\,\sin(s_\mathrm{I}\,n_0)&=A_\mathrm{II}\,\exp(\ri\,s_\mathrm{II}\,n_0)
+B_\mathrm{II}\,\exp(-\ri\,s_\mathrm{II}\,n_0).\\ \label{0.43}
A\,\sin[s_\mathrm{I}\,(n_0+1)]&=A_\mathrm{II}\,\exp[\ri\,s_\mathrm{II}\,(n_0+1)]
+B_\mathrm{II}\,\exp[-\ri\,s_\mathrm{II}\,(n_0+1)].
\end{align}
If the region II is forbidden, of $A_\mathrm{II}$ and $B_\mathrm{II}$
the coefficient corresponding to a negative imaginary part for the wave number
should vanish. Choosing by convention that the imaginary part of $s_\mathrm{II}$
be nonnegative, it is seen that in that case $B_\mathrm{II}$ is vanishing,
so that the boundary conditions at $n=n_0$ become
\begin{align}\label{0.44}
A\,\sin(s_\mathrm{I}\,n_0)&=A_\mathrm{II}\,\exp(\ri\,s_\mathrm{II}\,n_0).\\ \label{0.45}
A\,\sin[s_\mathrm{I}\,(n_0+1)]&=A_\mathrm{II}\,\exp[\ri\,s_\mathrm{II}\,(n_0+1)].
\end{align}
The condition that these equations have nontrivial solutions for $A$ and
$A_\mathrm{II}$ is then
\begin{equation}\label{0.46}
\sin[s_\mathrm{I}\,(n_0+1)]=\exp(\ri\,s_\mathrm{II})\,\sin(s_\mathrm{I}\,n_0).
\end{equation}
If the region II is forbidden, the wave-function decays exponentially in that region.
Such wave-functions correspond to bound-states. One notes that the region II
is forbidden, either if the energy is less than some value
($U_\mathrm{II}$), or when it is larger than some value ($U_\mathrm{II}+2\,K_0$),
unlike the case of continuum that bound-states occur only if the energy is less than
a certain value. It is also seen that if $K_0$ tends to infinity (which corresponds to
$\ell\to 0$, the continuum), the new possibility for bound-states is lost and
one recovers the continuum results. One also notes that of the two possibilities
resulting in the region II being forbidden, only one can be realized. The reason is that
the whole lattice cannot be forbidden. If $U_\mathrm{I}$ is smaller than $U_\mathrm{II}$,
the so called potential well, then the only way that the region II be forbidden is
that $E$ be smaller than $U_\mathrm{II}$. If $U_\mathrm{I}$ is larger than $U_\mathrm{II}$,
the so called potential barrier, then the only way that the region II be forbidden is
that $E$ be larger than $(U_\mathrm{II}+2\,K_0)$.

A shift in the potential energy is equivalent to a shift in the energy.
So without loss of generality one can put
\begin{equation}\label{0.47}
U_\mathrm{II}=0.
\end{equation}
There remains then, only $U_\mathrm{I}$. It is denoted by $\pm U_0$,
where $U_0$ is positive. The minus (plus) sign, then corresponds to
a potential well (barrier).
\subsection{The potential well}
There are two possible regions for the energy:
\begin{equation}\label{0.48}
\begin{cases}
\mbox{continuous spectrum},& 0\leq E\leq 2\,K_0\\
\mbox{discrete spectrum},& -U_0<E<\min(0,2\,K_0-U_0)
\end{cases}.
\end{equation}
For the discrete case, the energies are determined through (\ref{0.46}).
One has
\begin{align}\label{0.49}
s_\mathrm{II}&=\ri\,\cosh^{-1}\left(1-\frac{E}{K_0}\right).\\ \label{0.50}
s_\mathrm{I}&=\cos^{-1}\left(1-\frac{E+U_0}{K_0}\right).
\end{align}
If
\begin{equation}\label{0.51}
U_0>2\,K_0,
\end{equation}
then there is a gap between the band of continuous spectrum, and
the possible band of the discrete spectrum.
\subsection{The potential barrier}
There are two possible regions for the energy:
\begin{equation}\label{0.52}
\begin{cases}
\mbox{continuous spectrum},& 0\leq E\leq 2\,K_0\\
\mbox{discrete spectrum},& \max(2\,K_0,U_0)<E<2\,K_0+U_0
\end{cases}.
\end{equation}
For the discrete case, the energies are determined through (\ref{0.46}).
One has
\begin{align}\label{0.53}
s_\mathrm{II}&=\pi+\ri\,\cosh^{-1}\left(\frac{E}{K_0}-1\right).\\ \label{0.54}
s_\mathrm{I}&=\cos^{-1}\left(1-\frac{E-U_0}{K_0}\right).
\end{align}
Again, there is a gap between the band of continuous spectrum, and
the possible band of the discrete spectrum, if (\ref{0.51})
is satisfied.

The number of the discrete energies can be studied as follows.
Changing $U_0$, while other quantities are kept constant,
results in a change of the discrete energies and their corresponding
eigenvectors. However, the derivative of the energies with respect to
$U_0$ does not contain the derivative of the eigenvector:
\begin{align}\label{0.55}
\frac{\rd E}{\rd U_0}&=\frac{\rd\langle\psi|H|\psi\rangle}{\rd U_0},\nonumber\\
&=\left\langle\psi\left|\frac{\rd H}{\rd U_0}\right|\psi\right\rangle,\nonumber\\
&=\sum_{n=1}^{n_0}\langle\psi|n\rangle\langle n|\psi\rangle,
\end{align}
where $|\psi\rangle$ is the normalized eigenvector of $H$ corresponding to
the eigenvalue $E$ in the discrete spectrum. (One notes that for the discrete
spectrum, the eigenvalues are actually normalizable.) It is then seen that
the derivative of $E$ with respect to $U_0$ is positive. Hence $E$ is
increasing with respect to $U_0$. A discrete energy is larger than $(2\,K_0)$.
Increasing $U_0$ makes it larger, so such an energy cannot move into
the continuous spectrum, as $U_0$ is increased.
This shows that a discrete energy level is not {\em lost}, when $U_0$ is increased.
However, it could be possible that an additional eigenvalue be moved from
the continuous spectrum into the discrete spectrum, as $U_0$ is increased.
So the number of discrete energies is a nondecreasing function of $U_0$.
The introduction of a new eigenvalue in the discrete spectrum, however,
is only possible when there is no gap between the band of continuous spectrum
and the possible band of the discrete spectrum. So when $U_0$ exceeds $(2\,K_0)$,
so that the gap develops, the number of eigenvalues in the discrete spectrum
remains constant. The maximum number of the discrete energies is equal
to the number of discrete energies for very large $U_0$. For $U_0\to\infty$,
the imaginary part of $s_\mathrm{II}$ tends to infinity, so that
the quantization condition (\ref{0.46}) takes the simple form
\begin{equation}\label{0.56}
\sin[s_\mathrm{I}\,(n_0+1)]=0,
\end{equation}
the solutions to which are
\begin{equation}\label{0.57}
s_\mathrm{I}=\frac{k\,\pi}{n_0+1},\qquad1\leq k\leq n_0,
\end{equation}
where $k$ is in integer. So the maximum number of the discrete energies
is $n_0$. The eigenvalues corresponding to (\ref{0.57}), are of course
\begin{equation}\label{0.58}
\frac{E_k}{K_0}=1+\frac{U_0}{K_0}-\cos\frac{k\,\pi}{n_0+1}.
\end{equation}

A similar argument can be used to find values that when $U_0$ exceeds them
a new eigenvalue appears in the discrete spectrum. The argument is that such an
eigenvalue should start from the top of the band of continuous spectrum, which
is $(2\,K_0)$, when generation of new discrete eigenvalues are possible. So
a new eigenvalue in the discrete spectrum appears as $U_0$ exceeds a value
for which
\begin{align}\label{0.59}
s_\mathrm{II}&=\pi.\\ \label{0.60}
s_\mathrm{I}&=\cos^{-1}\left(\frac{U_0}{K_0}-1\right).
\end{align}
For these, the quantization condition (\ref{0.46}) becomes
\begin{equation}\label{0.61}
\sin[s_\mathrm{I}\,(n_0+1)]=-\sin(s_\mathrm{I}\,n_0),
\end{equation}
which results in
\begin{equation}\label{0.62}
s_\mathrm{I}=\pi-\frac{(2\,k-1)\,\pi}{2\,n_0+1},\qquad 1\leq k\leq n_0,
\end{equation}
where $k$ is integer. The corresponding value for $U_0$ would be
\begin{equation}\label{0.63}
\frac{U_0}{K_0}=1-\cos\frac{(2\,k-1)\,\pi}{2\,n_0+1}.
\end{equation}
In particular, the discrete spectrum is empty if
\begin{equation}\label{0.64}
\frac{U_0}{K_0}<1-\cos\frac{\pi}{2\,n_0+1}.
\end{equation}

Of course similar arguments hold for the potential well as well.
\subsection{Examples}
Consider the potential barrier. The problem is characterized by the dimensionless
values $(U_0/K_0)$, and $n_0$. As an example, consider
\begin{equation}\label{0.65}
n_0=6.
\end{equation}
The value of $(U_0/K_0)$ determines whether there is a gap between
the band corresponding to the discrete energies and the band of
continuous energies, or not.\\
\textbf{Small barrier.} Consider
\begin{equation}\label{0.66}
\frac{U_0}{K_0}=1.5.
\end{equation}
Here there is no gap between the band of discrete energies, $(E/K_0)\in[2,3.5]$,
and the band of continuous energies, $(E/K_0)\in[0,2]$. As
\begin{equation}\label{0.67}
1-\cos\frac{7\,\pi}{13}<1-\frac{U_0}{K_0}<1-\cos\frac{9\,\pi}{13},
\end{equation}
it is expected form (\ref{0.63}) that there are $4$ discrete energies.
The discrete energies, obtained by the quantization condition (\ref{0.46})
and checked by approximate methods, such as the Rayleigh-Ritz perturbation method
in the position basis \cite{Merz}, turn out to be
\begin{alignat}{2}\label{0.68}
\frac{E_1}{K_0}&=2.33248,&\qquad\frac{E_2}{K_0}&=2.76619,\nonumber\\
\frac{E_3}{K_0}&=3.14779,&\qquad\frac{E_4}{K_0}&=3.40786.
\end{alignat}
\textbf{Large barrier.} Consider
\begin{equation}\label{0.69}
\frac{U_0}{K_0}=2.5.
\end{equation}
Here there is a gap between the band of discrete energies, $(E/K_0)\in[2.5,4.5]$,
and the band of continuous energies, $(E/K_0)\in[0,2]$. The discrete energies,
obtained by the quantization condition (\ref{0.46}) and checked by approximate methods,
such as the Rayleigh-Ritz perturbation method \cite{Merz}, turn out to be
\begin{alignat}{2}\label{0.70}
\frac{E_1}{K_0}&=2.60654,&\qquad\frac{E_2}{K_0}&=2.89816,\nonumber\\
\frac{E_3}{K_0}&=3.30698,&\qquad\frac{E_4}{K_0}&=3.74878,\nonumber\\
\frac{E_5}{K_0}&=4.13895,&\qquad\frac{E_6}{K_0}&=4.40548.
\end{alignat}
\textbf{Very large barrier.} By this, it is meant that
\begin{equation}\label{0.71}
\frac{U_0}{K_0}\gg 1.
\end{equation}
The discrete energies turn out to be
\begin{alignat}{2}\label{0.72}
\frac{E_1}{K_0}&=1+\frac{U_0}{K_0}-0.90097,&\qquad\frac{E_2}{K_0}&=1+\frac{U_0}{K_0}-0.62349,\nonumber\\
\frac{E_3}{K_0}&=1+\frac{U_0}{K_0}-0.22252,&\qquad\frac{E_4}{K_0}&=1+\frac{U_0}{K_0}+0.22252,\nonumber\\
\frac{E_5}{K_0}&=1+\frac{U_0}{K_0}+0.62349,&\qquad\frac{E_6}{K_0}&=1+\frac{U_0}{K_0}+0.90097.
\end{alignat}
In fact,
\begin{equation}\label{0.73}
\frac{E_k}{K_0}=1+\frac{U_0}{K_0}-\cos\frac{k\,\pi}{7},
\end{equation}
which is the same as (\ref{0.58}).
\section{A toy model one-dimensional lattice}
In this section, a Kronig-Penney model (on a one-dimensional line)
is tailored so that it behaves essentially like a lattice model,
to be specific, like something with one single (compact) band for
the kinetic energy. The model considered here consists of a particle
moving on a line under the influence of a periodic (infinite) array
of attractive Dirac delta potentials. It is seen that by adjusting
the parameters, one can make the distance between the lowest
and the next energy band large, while keeping the position and
the width of the band finite, so that essentially there remains only
only energy band. This would be like the case of a free particle
moving on a one dimensional infinite lattice.

Consider a particle of mass $m$ moving on a line and experiencing
a potential energy $V$, with
\begin{equation}\label{0.74}
V(x)=V_\mathrm{r}-V_0\,\ell\,\sum_j\delta(x-j\,\ell),
\end{equation}
where $V_\mathrm{r}$, $V_0$, and $\ell$ are real constants, $V_0$ and $\ell$ being positive.
The eigenvalue problem for the Hamiltonian would be
\begin{equation}\label{0.75}
-\frac{\hbar^2}{2\,m}\,\frac{\partial^2\,\psi(x)}{\partial\, x^2}+V(x)\,\psi(x)=E\,\psi(x).
\end{equation}
The solution to the above in the interval $[-\ell,\ell]$ is
represented as
\begin{equation}\label{0.76}
\psi(x)=\begin{cases}C\,\exp(\kappa\,x)+D\,\exp(-\kappa\,x),& -\ell<0<x\\
C'\,\exp(\kappa\,x)+D'\,\exp(-\kappa\,x),& 0<x<\ell
\end{cases}
\end{equation}
where
\begin{align}\label{0.77}
E=-\frac{\hbar^2\,\kappa^2}{2\,m}+V_\mathrm{r},
\end{align}
The continuity conditions at $x=0$ read
\begin{align}\label{0.78}
C'+D'&=C+D.\\ \label{0.79}
\frac{\hbar^2\,\kappa}{2\,m}\,(C'-D')&=\frac{\hbar^2\,\kappa}{2\,m}\,(C-D)-V_0\,\ell\,(C+D).
\end{align}
Also, as the Hamiltonian commutes with translations by the amount $\ell$,
one can seek eigenvectors of the Hamiltonian which are at the same time
eigenvectors of the translation by the amount $\ell$ (or any integer
multiple of that). For such eigenvectors, one has
\begin{equation}\label{0.80}
\psi(x+\ell)=\exp(\ri\,\alpha)\,\psi(x),
\end{equation}
where $\alpha$ is a real constant. (In the terminology of
conventional continuum quantum mechanics, such a wave-function
$\psi$ is called a Bloch function.) So $\psi(x)\,\exp(-\ri\,\alpha\,x/\ell)$
is periodic in $x$ with the period $\ell$, resulting in
\begin{align}\label{0.81}
C'\,\exp(\kappa\,\ell)&=C\,\exp(\ri\,\alpha).\\ \label{0.82}
D'\,\exp(-\kappa\,\ell)&=D\,\exp(\ri\,\alpha).
\end{align}
Using these to eliminate $C'$ and $D'$, and defining
\begin{align}\label{0.83}
\frac{m\,V_0\,\ell^2}{\hbar^2}&=:\upsilon,\\ \label{0.84}
\kappa\,\ell&=:\beta,
\end{align}
by which
\begin{align}\label{0.85}
E=-V_0\,\frac{\beta^2}{2\,\upsilon}+V_\mathrm{r},
\end{align}
one arrives at
\begin{align}\label{0.86}
\begin{bmatrix}\exp(\beta)&0\\0&\exp(-\beta)\end{bmatrix}\,\begin{bmatrix}
1-(\upsilon/\beta)&-(\upsilon/\beta)\\
(\upsilon/\beta)&1+(\upsilon/\beta)\end{bmatrix}\,\begin{bmatrix}C\\D\end{bmatrix}
&=\exp(\ri\,\alpha)\,\begin{bmatrix}C\\D\end{bmatrix}.
\end{align}
The eigenvalue problem (\ref{0.86}) results in
\begin{equation}\label{0.87}
\exp(\beta)\,\left(1-\frac{\upsilon}{\beta}\right)
+\exp(-\beta)\,\left(1+\frac{\upsilon}{\beta}\right)=2\,\cos\alpha,
\end{equation}
or equivalently
\begin{equation}\label{0.88}
\cosh\beta-\frac{\upsilon}{\beta}\,\sinh\beta=\cos\alpha.
\end{equation}
$\beta$ could be real or imaginary. (\ref{0.85}) shows that the eigenvalues of $H$
which are less than $V_\mathrm{r}$, correspond to real values of $\beta$.
Obviously, for $\upsilon\to\infty$, the solution of above is
$\beta=\upsilon$. Based on this, one can develop perturbative solutions
for (\ref{0.88}). For large values of $\upsilon$ one takes
\begin{align}\label{0.89}
\beta&=\upsilon+\gamma,
\end{align}
with $|\gamma|\ll\upsilon$, so that
\begin{align}\label{0.90}
\exp(-\upsilon)+\left[\frac{\sinh\upsilon}{\upsilon}-\exp(-\upsilon)\right]\,\gamma+\cdots&=
\cos\alpha.
\end{align}
Solving for $\gamma$, one arrive at
\begin{equation}\label{0.91}
\beta=\upsilon\,[1+2\,\exp(-\upsilon)\,\cos\alpha+\cdots],
\end{equation}
from which
\begin{equation}\label{0.92}
E=-V_0\,\frac{\upsilon}{2}\,[1+4\,\exp(-\upsilon)\,\cos\alpha]+V_\mathrm{r}+\cdots.
\end{equation}
So, as $\cos\alpha$ takes values in the interval $[-1,1]$, the allowed energies
form a band of width $W$, separated by a gap of at least $\Delta$ from
the next band (the energies larger than $V_\mathrm{r}$, which correspond to imaginary
values for $\beta$), where
\begin{align}\label{0.93}
W&=4\,V_0\,\upsilon\,\exp(-\upsilon).\\ \label{0.94}
\Delta &=V_0\,\frac{\upsilon}{2}.
\end{align}
Clearly,
\begin{equation}\label{0.95}
\lim_{\upsilon\to\infty}\frac{W}{\Delta}=0,
\end{equation}
meaning that while the width of band can be kept finite, the gap can
be set so large that the possible energies are effectively bounded.
One can tune $V_0$ and $V_\mathrm{r}$ so that the width of the energy band
remains finite and the position of the energy band remains finite
as well (as $\upsilon$ tends to infinity):
\begin{align}\label{0.96}
V_0&=\frac{W\,\exp(\upsilon)}{4\,\upsilon},\\ \label{0.97}
V_\mathrm{r}&=\frac{W\,\exp(\upsilon)}{8},
\end{align}
for a finite and fixed value of $W$. With these, in the limit that
$\upsilon$ tends to infinity, the set of the finite eigenvalues of
the Hamiltonian becomes a single energy band. This energy band
could be regarded as the values of the kinetic energy for
a system on a lattice. One notes that this energy band is actually
the sum of a potential energy and a kinetic energy for the original
system on the continuum, but the effective spectrum of the total
Hamiltonian on the continuum, which is invariant under space translation
by the size of the lattice, is the same as the spectrum of kinetic energy
of a particle moving on a lattice.

\vspace{\baselineskip}
\noindent\textbf{Acknowledgement}: This work is
supported by the Research Council of Alzahra University.

\end{document}